\documentclass[amsmath,amssymb,reprint,floatfix,superscriptaddress,showkeys]{revtex4-1} 
\usepackage[english]{babel}
\usepackage[utf8]{inputenc}

\usepackage{graphicx}

\usepackage[version=3]{mhchem}
\usepackage{units}
\usepackage{textcomp}
\usepackage{hyperref}
\usepackage[capitalize]{cleveref}

\begin{document}

\title{Variable resistor made by repeated steps of epitaxial deposition and lithographic structuring of oxide layers by using wet chemical etchants}


\author{Dieter~Weber}
\email[]{d.weber@fz-juelich.de}

\author{Róza~Vőfély}
\altaffiliation{Currently Trinity College, University of Cambridge, Cambridge, UK}

\author{Yuehua~Chen}
\altaffiliation{Currently Xi'an Jiaotong University, Xi'an, China}

\affiliation{Forschungszentrum Jülich, Peter Grünberg Institute PGI-5, 52425 Jülich, Germany}

\author{Yulia~Mourzina}
\affiliation{Forschungszentrum Jülich, Peter Grünberg Institute PGI-8/Institute of Complex Systems ICS-8, 52425 Jülich, Germany}

\author{Ulrich~Poppe}
\affiliation{Forschungszentrum Jülich, Peter Grünberg Institute PGI-5, 52425 Jülich, Germany}

\date{19 January 2013}

\begin{abstract}
Variable resistors were constructed from epitaxial \ce{SrRuO3} (SRO), \ce{La_{0.67}Sr_{0.33}MnO3} (LSMO) and \ce{SrTiO3} layers with perovskite crystal structure. Each layer was patterned separately by lithographic methods. Optimized wet chemical etchants and several polishing steps in organic solvents allowed good epitaxy of subsequent layers, comparable to epitaxy on pristine substrates. Periodate as the oxidizing agent for SRO and iodide with ascorbic acid as the reducing agents for LSMO were used to attack these chemically resistant oxides. The final devices changed their conductance in a similar manner to previously described variable resistors that were defined with shadow masks.

\vspace{3mm}
\emph{Thin Solid Films} (2013), \url{http://dx.doi.org/10.1016/j.tsf.2012.11.118}
\end{abstract}


\maketitle

\section{Introduction}
\label{intro}
Ionic transport on the nanoscale at very high electrical field strength in the \unitfrac{MV}{cm} range is crucial for the functioning of redox memory and electrochemical metalization cells~\cite{Waser2007}. These two alternative solid state computer memory concepts promise properties that could in the future surpass the established Flash EEPROM~\cite{ITRS2011}. The very high electrical field contributes a significant part of the activation energy of ionic jump processes in these devices, which leads to an exponential increase of transport speed with applied voltage~\cite{Lamb-field-acceleration,Meyer-CMOX,Strukov2009}. Localized heating caused by high local current density can also contribute to accelerated ionic transport~\cite{ADFM:ADFM201101117}. We showed in a previous work~\cite{weber:056101} that an all-oxide epitaxial crossbar structure with a \ce{SrTiO3} (STO) dielectric sandwiched between Sr-doped \ce{La2CuO4} electrodes has a highly systematic switching behavior of the in-plane electrode 
conductivity as a function of applied 
current and write time, which could help to investigate oxygen transport in the high-field regime. The device that we used in the previous work is, however, so large that the parasitic voltage drop inside the conducting electrodes leads to localized writing where the field strength is highest. The thickness of the device structures is, furthermore, uneven due to the shadowing effect of the masks. This makes a quantitative interpretation of the results very difficult. Analysis of the device's scaling behavior shows that smaller devices should be affected by localized writing to a lesser degree. Therefore it is desirable to not use shadow masks but to create such structures by using lithographic methods that allow dimensions and precise alignment down to the µm range.

Doped \ce{La2CuO4} for the electrodes is not an ideal choice for patterning with photolithography because this material is hygroscopic and nanometer-thick films degrade within hours in contact with moist air and very quickly in contact with water. The developers for common photoresists are alkaline aqueous solutions, and deionized water is used for rinsing after development. This means that electrodes and alignment markers made from this material do not survive repeated patterning steps with standard photolithography. It is therefore desirable to choose different materials for the conducting electrodes that are stable at least in moist air, water and alkaline aqueous media. Two stable oxides with good conductivity are \ce{SrRuO3} (SRO) and \ce{La_{0.67}Sr_{0.33}MnO3} (LSMO). Similar to \ce{La2CuO4}, LSMO changes its resistivity as a function of oxygen content which should make it suitable as an electrode material~\cite{wilson:4971}.

Ion beam milling with argon is commonly used to pattern such oxide thin films, but it has three disadvantages. First, the thickness of etched films with only photoresist as the mask is limited to about 50\,nm to 100\,nm because of the fast degradation of the resist under ion bombardment. Second, the milling rate of all oxides is similar so that only mass spectrometer data or precise timing can be used to determine the end point. And third, ion bombardment amorphizes the surface and can lead to a stoichiometry change because of preferential sputtering~\cite{Otte2000498,faley:2138,jia:3635}. This makes subsequent epitaxy difficult and leads to rough layers, misaligned crystal grains and precipitates in following layers. Argon ion beam milling is therefore only acceptable if the layer quality on top of etched areas is not a concern.

In contrast to argon ion beam milling, wet chemical etchants can be selective for a specific material without attacking the photoresist and other materials present on the sample. They can have very high etch rates and also allow underetching because of their undirected attack. If they are properly chosen for a given material combination, they leave a clean surface without damage to the underlying crystal structure~\cite{faley:2138,jia:3635}. Selective wet chemical etching is therefore advantageous to create complex epitaxial heterostructures with repetitive patterning and deposition steps. A proper combination of directional, unselective ion beam etching with selective, isotropic chemical etching provides a wide range of possibilities to build complex three-dimensional heterostructures. Chemical etchants have to be developed and optimized individually for each material and material combination. Much less information on chemical etching of common functional oxides is available compared to other materials 
used in the semiconductor industry. The amount of work required until a viable manufacturing procedure is found can therefore be significant.

For LSMO, concentrated hydrochloric acid~\cite{stroud:7189}, \ce{HCl} with \ce{KI} in water~\cite{Lee-LSMO_etch}, ammonia-buffered hydrofluoric acid (BHF)~\cite{springerlink:10.1007/s11664-007-0343-x} and citric acid solution at elevated temperature~\cite{5706349} are known etchants. All these etchants rely on the reduction of insoluble \ce{Mn(IV)} in LSMO to more soluble \ce{Mn(II)}. Mn(IV) can oxidize \ce{Cl-} to \ce{Cl2} according to \ce{MnO2 + 4 H3O+ + 2 Cl- -> Mn^{2+} + 6 H2O + Cl2}~\cite{Wiberg1995}. If Mn(IV) is not reduced, insoluble \ce{MnO2} precipitates are left behind. 10\,\% HF, for example, does not etch LSMO~\cite{Biasotti2009839}, in contrast to BHF containing ammonia, which can act as a reducing agent~\cite{Wiberg1995}. The reaction is perhaps catalyzed by \ce{F-} through an intermediate species, such as \ce{[MnF6]^{2-}}, that in turn oxidizes ammonia~\cite{Hoppe1957}.

For SRO, only a solution of ozone in water~\cite{SRO-Ozone-Patent} is known as a wet etchant. The ruthenium in SRO has the oxidation number IV as in \ce{RuO2}. This compound is insoluble in acids and bases and is responsible for the chemical stability of SRO. Oxidation of \ce{RuO2} to volatile \ce{RuO4} with \ce{NaIO4} and other oxidants dissolved in water is documented by Griffith~\cite{griffith2010ruthenium}. This suggests that such oxidants can also etch SRO according to \cref{eq:SRO-etch} by turning the insoluble \ce{Ru(IV)} into \ce{RuO4}. The remaining strontium oxide is then easily dissolved in the aqueous solution. 

\begin{equation}
\begin{split}
 \ce{SrRuO3(s) } & \ce{+ 2 H3O+ + 2 IO4- ->} \\
  & \ce{Sr^2+(aq) + RuO4(aq) + 3 H2O + 2 IO3-} \label{eq:SRO-etch}
\end{split}
\end{equation}

Pourbaix diagrams~\cite{Pourbaix1974} of all elements in a sample together with Pourbaix diagrams of potential etchants can help to find a good composition and pH range.

\section{Experimental details}
\begin{figure}
 \centering
 \includegraphics[width=8cm]{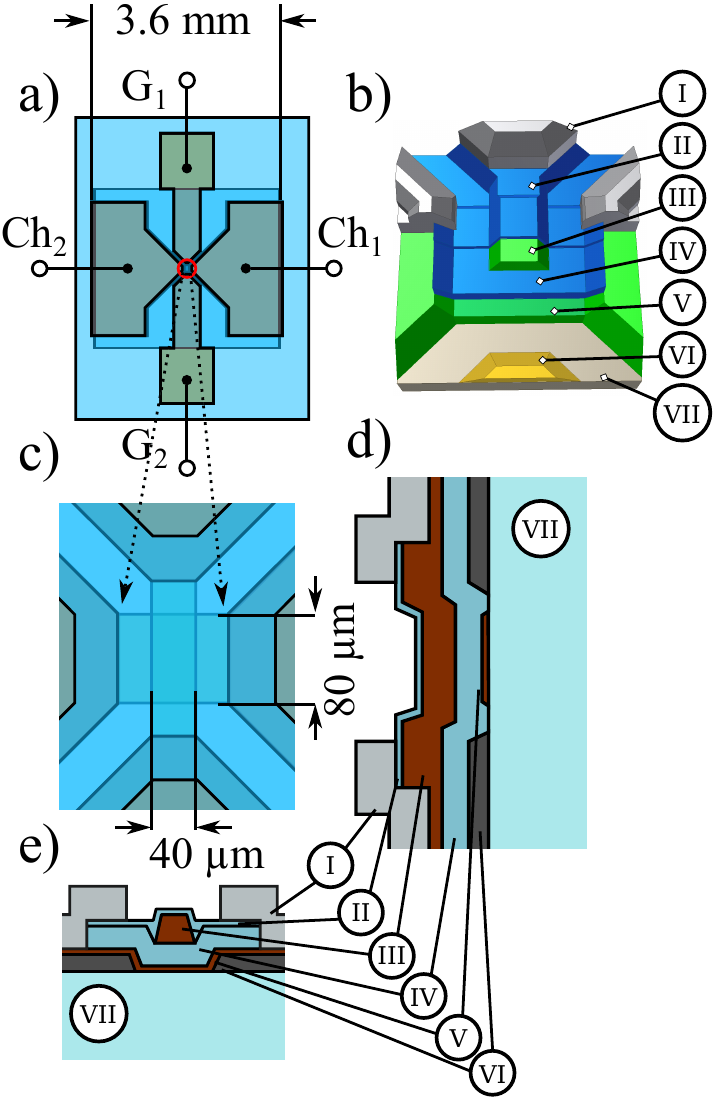}
 \caption{Diagram of the device structure. A plane view of the entire device is shown in a). b) is a 3D view of the crossover area. A plane view magnification of the crossover area is shown in c). d) and e) are cross-sectional views of the crossover area. A thin LSMO bridge (V) at the bottom forms the channel and is contacted on both ends by SRO (VI) (terminals $Ch_1, Ch_2$). The STO dielectric (IV) covers the cross-section area completely and is topped with the gate electrode (III) in the form of another bridge orthogonal to the channel with terminals $G_1, G_2$. The device is then covered with a thin STO passivation layer (II) protecting against atmospheric influences. The silver layer (I) contacts the channel and the gate through holes in the passivation and the dielectric layer.}
 \label{fig:sketch}
\end{figure}

Here we present a miniaturized version of the previously shown device~\cite{weber:056101}. This new version was manufactured by lithographic methods instead of shadow masks and uses SRO and LSMO instead of doped \ce{La2CuO4} for the conducting electrodes. The device structure is shown schematically in \cref{fig:sketch}. The materials, thicknesses and heater temperatures during deposition of the layers that compose the device structure are listed in \cref{tab:layers}. The channel has a width of 80\,µm and the gate has a width of 40\,µm. These dimensions were chosen as a compromise. On the one hand, the problems with parasitic voltage drops are reduced by miniaturization, but on the other hand, the resistance between channel and gate through the dielectric has to be low enough to allow measurements of the dielectric's characteristic curve with the given instruments.

\begin{table}
\centering
\caption{Material, heater temperature during deposition and thickness of the device layers}
\label{tab:layers}
\begin{tabular}{llrr@{$\pm$}l}
\toprule
\textbf{Layer} & \textbf{Material} & \multicolumn{1}{p{1.6cm}}{\textbf{Heater/\textcelsius}} & \multicolumn{2}{l}{\textbf{Thickness/nm}}\\
\hline
Contacts       & SRO    & 700 & 27&1\\
Channel        & LSMO   & 900 &  9&0.5\\
Dielectric     & STO    & 800 & 42&2\\
Gate           & LSMO   & 900 & 67&2\\
Passivation    & STO    & 800 &  6&0.5\\
Metal contacts & Silver & RT  &\multicolumn{2}{l}{$\approx100$} \\
\botrule
\end{tabular}
\end{table}

All the layers were deposited in a high-pressure oxygen sputtering system~\cite{Poppe-sputter} on BHF-terminated STO(100)k substrates from Crystec, Berlin. The deposition temperatures are to be understood as the heater temperature. Actual sample temperatures vary with the gas pressure. 

The samples were first heated to 950\,\textcelsius\ heater temperature under vacuum with a background value lower than $\unit[4\cdot10^{-6}]{mbar}$. Then they were cooled to deposition temperature at an oxygen pressure of 100\,mbar. Subsequently, the pressure was lowered to 3\,mbar and the target was pre-sputtered for 5 minutes. The deposition rate for SRO was about \unitfrac[22]{nm}{h}, for LSMO \unitfrac[55]{nm}{h} and for STO \unitfrac[12.5]{nm}{h}. LSMO layers were heated to 950\,\textcelsius\ after deposition, the pressure was increased to 500\,mbar \ce{O2} and the sample was annealed for 1\,h, during which it cooled to about 850\,\textcelsius\ due to limited heater power and increased cooling by the gas atmosphere. Other layers were directly cooled down at a pressure of 500\,mbar.

\begin{table*}
\centering
\caption{Results of experiments to choose suitable etchants for LSMO. Solvent is water if not otherwise stated. The etch rate is to be understood as an approximate value because at least 15\,s over-etching is necessary to remove the last residuals of a layer. The etchant printed in bold face was chosen for the actual device.}
\label{tab:etchLSMO}
\begin{tabular}{p{8cm}rp{5cm}}
\toprule
\textbf{Etchant} & \textbf{Etch rate} & \textbf{Remarks}\\
\hline
37\,wt-\% HCl & \unitfrac[7]{nm}{s} & Precipitates on surface\\
15\,wt-\% HCl & Very slow & Brown haze, dirty surface \\
Saturated citric acid at 85\,\textcelsius & \unitfrac[0.2]{nm}{s} & Residuals \\
0.37\,\% HCl, \unitfrac[5]{mol}{l} \ce{KI} & \unitfrac[6]{nm}{s} & Solution turns brown in air (\ce{I-} oxidized to \ce{I3-} by \ce{O2})\\
3.7\,\% HCl, \unitfrac[5]{mol}{l} \ce{KI} & \unitfrac[28]{nm}{s} & Solution turns brown in air\\
\textbf{0.37\,\% HCl, \unitfrac[5]{mol}{l} \ce{KI}, \unitfrac[0.1]{mol}{l} ascorbic acid} & \textbf{\unitfrac[6]{nm}{s}} & \textbf{Solution remains clear}\\
\unitfrac[1]{mol}{l} ascorbic acid, \unitfrac[0.01]{mol}{l} EDTA, \unitfrac[1.05]{mol}{l} \ce{NH3} (pH 4)& Very slow & Brown haze, poor surface \\
Saturated ascorbic acid, \unitfrac[0.01]{mol}{l} 18-crown-6 in ethanol & No effect & \\
1\,\% methanesulfonic acid, 1\,\%~glutaraldehyde, 1\,\%~\ce{H2O}, \unitfrac[0.01]{mol}{l}~18-crown-6 in ethanol & Very slow & Brown haze, poor surface \\                                                   
\botrule
\end{tabular}
\end{table*}

\begin{table}
\centering
\caption{Results of experiments to choose suitable etchants for SRO. Solvent is water if not otherwise stated. The etchant printed in bold face was chosen for the actual device.}
\label{tab:etchSRO}
\begin{tabular}{p{3.6cm}rp{2cm}}
\toprule
\textbf{Etchant} & \textbf{Etch rate} & \textbf{Remarks}\\
\hline
10.9\,\%~\ce{[Ce(NO3)6](NH4)2}, 4.25\,\%~\ce{HClO4} & $> \unitfrac[100]{nm}{s}$ & Particles or precipitates on surface \\
\unitfrac[0.4]{mol}{l} \ce{H5IO6} & \unitfrac[20]{nm}{s} \\
\textbf{\unitfrac[0.4]{mol}{l} \ce{NaIO4}} & \textbf{\unitfrac[2]{nm}{s}} \\
\unitfrac[0.4]{mol}{l} \ce{NaClO4} & No effect \\
\unitfrac[0.4]{mol}{l} \ce{NaClO3} & Very slowly \\%
\unitfrac[0.05]{mol}{l} \ce{H5IO6}, \unitfrac[0.01]{mol}{l} 18-crown-6 in ethanol & No effect \\
\botrule
\end{tabular}
\end{table}

Several etch and growth tests were carried out with numerous test samples of LSMO and SRO layers. The results are given in \cref{tab:etchLSMO} and \ref{tab:etchSRO}. Polishing the surface repeatedly with a cotton swab inside water or organic solvents helped to reduce the number and size of precipitates on the surface after etching, because the oxide crystal surface is mechanically much harder than the floccose and loosely attached precipitates. The precipitates seem to appear especially after treatment with strong acids. A possible explanation is that strong acids leach \ce{Sr^2+} from the STO surface according to \cref{eq:Sr-leach}, leaving \ce{TiO2} or the corresponding hydroxide behind.

\begin{equation}
 \ce{SrTiO3(s) + 2 H3O+ ->T[{low pH}] Sr^2+(aq) + 3 H2O + TiO2(s)} \label{eq:Sr-leach}
\end{equation}

LSMO layers that grew on an STO surface after concentrated HCl etching of LSMO had a higher roughness than LSMO layers after etching with dilute HCl and KI. This etchant with ascorbic acid as antioxidant to prevent the solution from turning brown upon contact with air was therefore chosen for LSMO. For SRO, \ce{NaIO4} solution was chosen because of its suitable etch rate and near-neutral pH.

AZ~5214E photoresist spin-coated at 4000\,rpm and the developer AZ~326~MIF were used for both etching (positive) and lift-off (negative) steps with the usual processing parameters.

After each chemical etching step, the samples were rinsed quickly in water and then rubbed immediately with a cotton swab inside ethanol or isopropanol in order to remove soft residuals or precipitates from the surface. This treatment also removed the photoresist. The samples were cleaned again with an ultrasonic treatment and by rubbing in both acetone and isopropanol before the next deposition step.

The STO dielectric and passivation layers were etched by argon ion beam milling under an incident angle of 45\textdegree\ while the sample was rotating. The etch rate was calibrated with a reference STO thin film deposited on \ce{NdGaO3} to detect the endpoint with a mass spectrometer. After the ion beam etching steps, the samples were cleaned by ultrasound in acetone and isopropanol and subsequently oxidized in a Tepla~300 plasma processor for 30~minutes at 300\,W power, 200\,sccm \ce{O2} and a pressure of about 0.7\,mbar in order to remove cross-linked resist residuals. The final metal contacts were created by lift-off of a sputtered silver layer.

The finished devices were measured by the same procedure and setup as described in our previous paper~\cite{weber:056101}. The current through the dielectric was applied to terminals $Ch_1, G_1$ and the voltage was measured on terminals $Ch_2, G_2$ (\cref{fig:sketch}). The positive current direction is defined as being from the channel to the gate.

\section{Results and discussion}

\subsection{Structural characterization}

\begin{figure*}
 \centering
 \includegraphics[width=18cm]{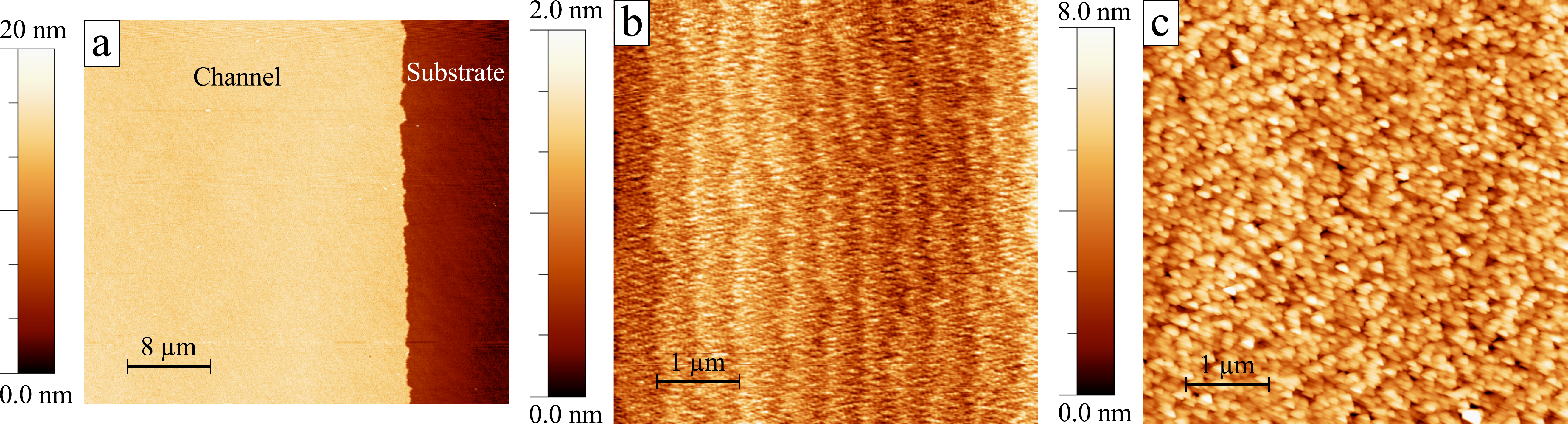}
 \caption{AFM images after deposition of the STO dielectric layer on chemically etched and ion beam milled surfaces. (a) shows the edge between LSMO channel and substrate. Both SRO contacts (not visible) and LSMO channel were etched chemically. Note the sharp rectangular step without any disturbances at the edge. (b) is a more detailed view on top of an etched part of the substrate. (c) shows STO growth on STO after argon ion beam milling for comparison. The surface is much rougher and has a grainy texture. A few scan lines where the tip left the surface are masked out.}
 \label{fig:STO-AFM}
\end{figure*}

The crossover area between channel and gate is critical for the operation of the device. This area should therefore be of very good quality with uniform layer thickness, smooth surface and without defects. \cref{fig:STO-AFM} (a) and (b) show AFM images after the deposition of the dielectric. The dielectric had a very smooth surface on top of the LSMO channel layer as well as the STO substrate without visible precipitates or misaligned grains. It covered the step between LSMO channel and substrate without any detectable disruption (\cref{fig:STO-AFM} (a)). \cref{fig:STO-AFM} (b) shows the excellent epitaxial growth of the STO dielectric layer on top of the STO substrate surface where SRO as well as LSMO were grown and subsequently etched away. For comparison, growing an STO dielectric layer on an STO substrate surface after LSMO etching with argon ion milling and subsequent annealing at about 900\,\textcelsius\ in 20\,mbar \ce{O2} for 1\,h (\cref{fig:STO-AFM} (c)) gave a much rougher surface with a grainy 
texture. This demonstrates how chemical etching is far superior to ion beam milling in order to structure buried epitaxial layers. 

\subsection{Electrical characterization}

\begin{figure}
 \centering
 \includegraphics[width=8cm]{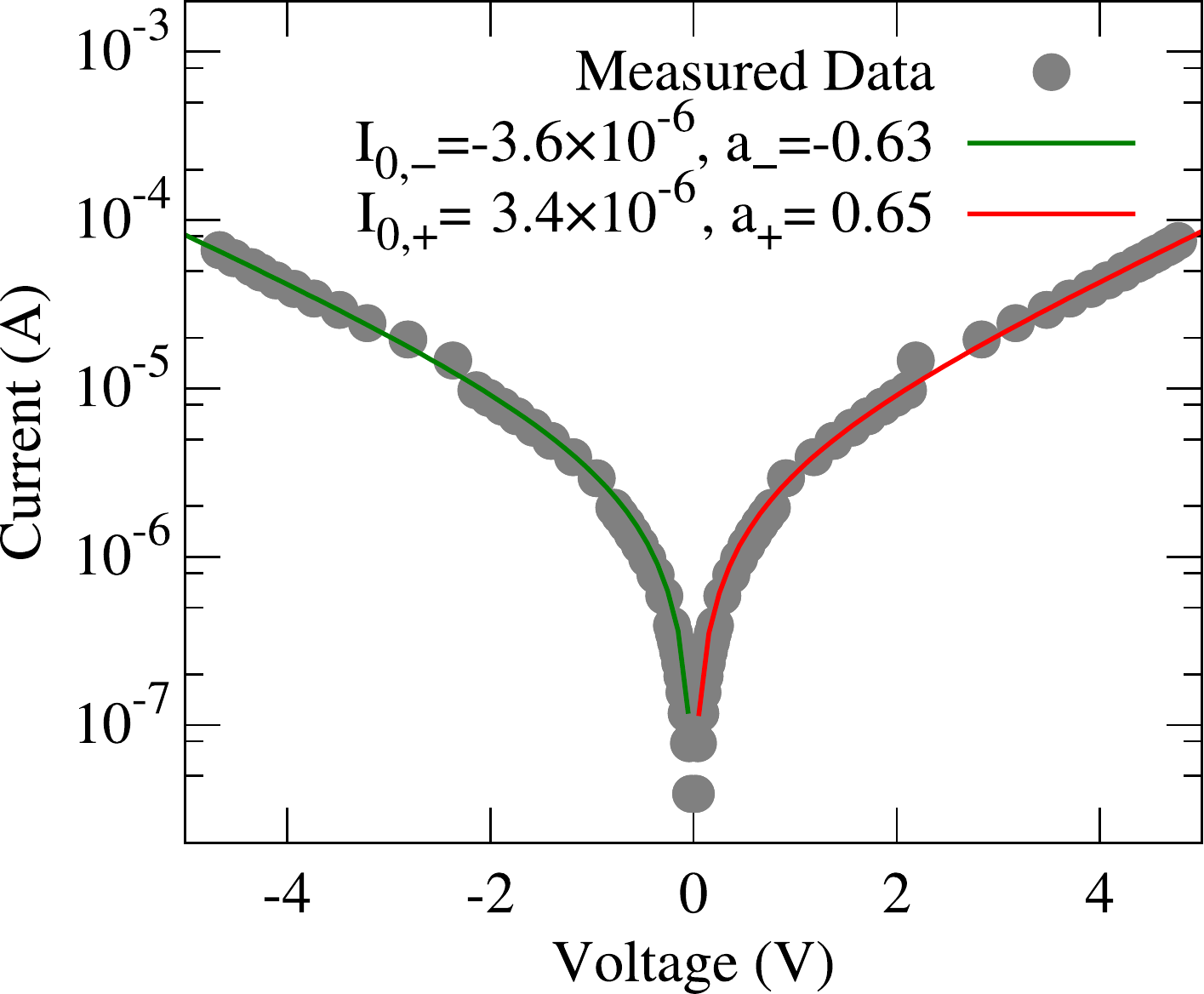}
 \caption{Current-voltage characteristic curve of the dielectric. The solid lines are a fit of the function $I_{el}(U) = I_{0,el} \left( \exp \left( a_{el} \times U \right) - 1 \right)$ for negative and positive values with the indicated fit parameters. The steps visible at around 2.5\,V are likely artifacts, because the complete dataset was combined from three subsequent measurements with different ranges.}
 \label{fig:IV}
\end{figure}

The electronic current $I_{el}$ through the dielectric as a function of voltage $U$ behaved like through two antiparallel series-connected diodes with large reverse current (\cref{fig:IV}) and could be described by the equation $I_{el} = I_{0,el} \left( \exp \left( a_{el} \times U \right) - 1 \right)$ for negative and positive currents with parameters $I_{0,el}$ and $a_{el}$. This behavior can be explained with each LSMO/STO junction in the LSMO/STO/LSMO stack behaving as a Schottky diode. A comparable behavior of STO/LSMO junctions was found by Yajima et al.~\cite{Yajima2011}.

\begin{figure}
 \centering
 \includegraphics[width=8cm]{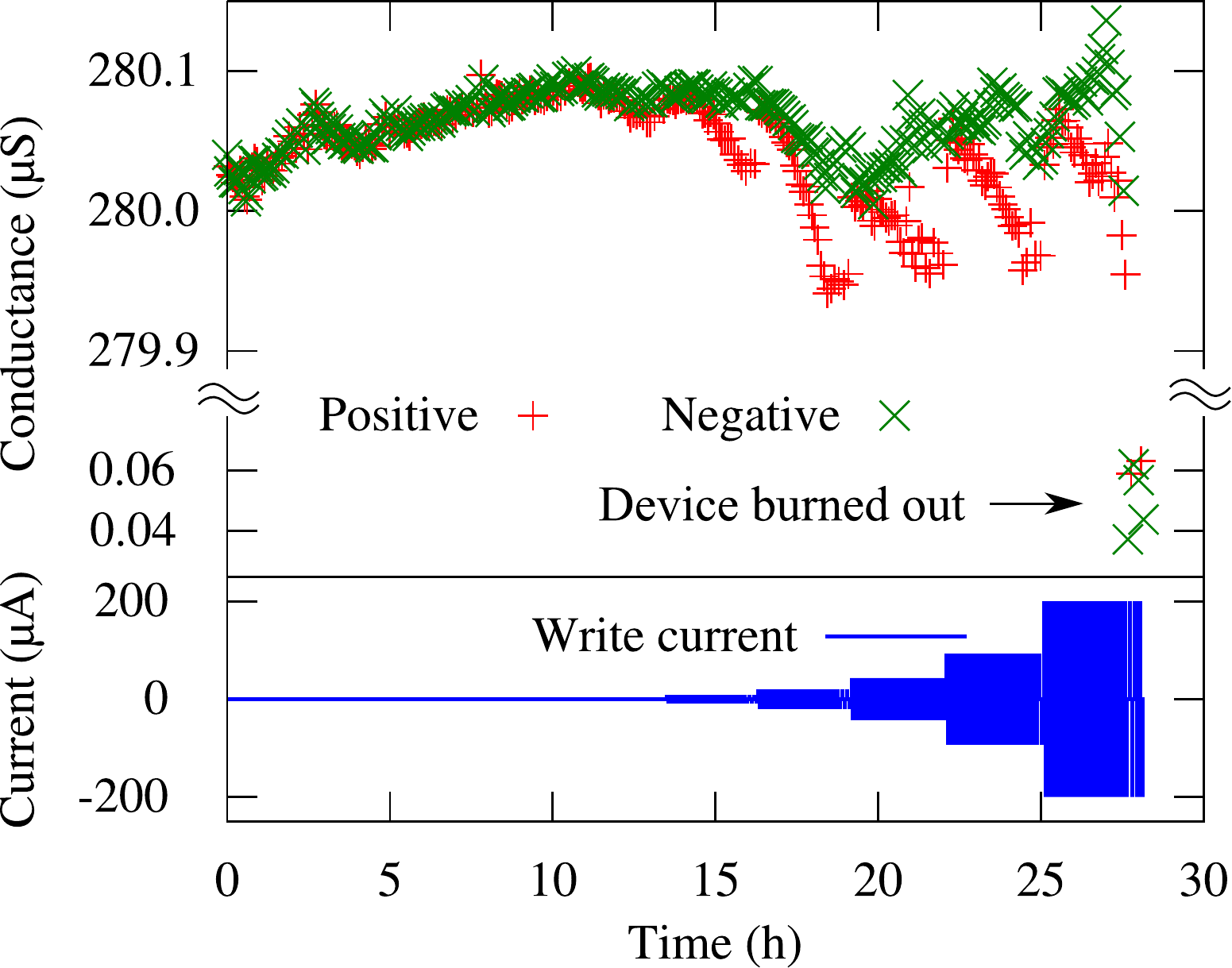}
 \caption{Channel conductance change induced by alternating write steps through the dielectric with increasing current and duration. The write time was increased from a minimum value of about 3\,s (lowest current value) or 3\,ms (highest current value) to 100\,s during each current step. The device burned out after around 28\,h, which led to an irreversible drop in the channel conductance.}
 \label{fig:RW}
\end{figure}

\begin{figure}
 \centering
 \includegraphics[width=6cm]{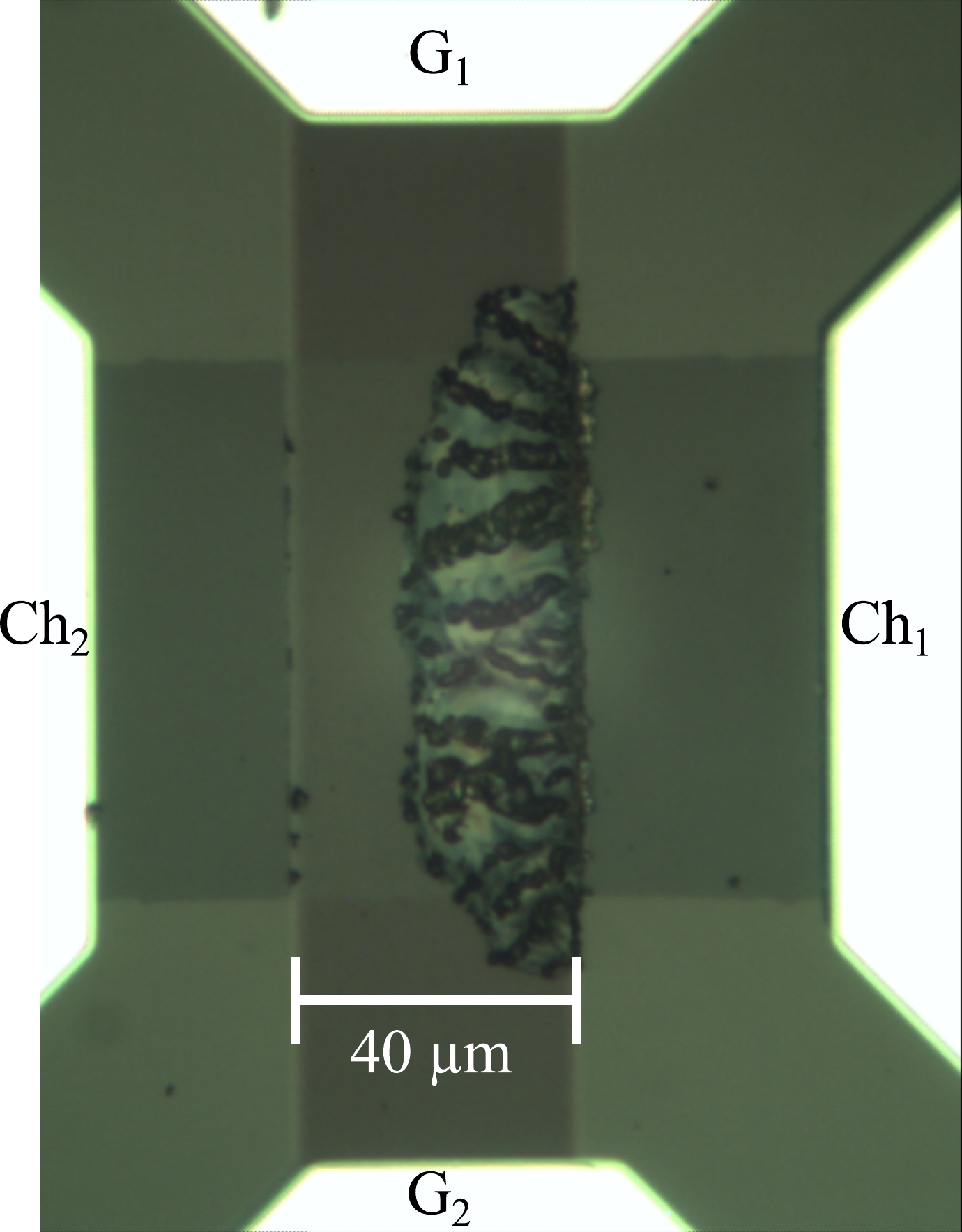}
 \caption{Optical micrograph of a device that burned out when writing with 200\,µA for 100\,s. The horizontal electrode is the gate and the vertical electrode is the channel. \cref{fig:sketch} shows a diagram of the structure for comparison.}
 \label{fig:burn}
\end{figure}

The conductance of the channel changed systematically with the applied current and duration (\cref{fig:RW}), but the relative change is about one order of magnitude smaller than that of our previous device~\cite{weber:056101} despite the higher maximum current density (now \unitfrac[6.25]{A}{cm²} instead of \unitfrac[2.4]{A}{cm²}). The current-voltage characteristic of the dielectric and the conductance of the gate did not change systematically as a function of the current direction within the measurable range. This excludes leakage paths through the dielectric causing the channel's systematic conductance change. The device burned out at the highest current of 200\,µA (\cref{fig:burn}) and at an approximate maximum write voltage of $\unit[6.5\pm0.5]{V}$. The voltage during writing was not measurable in our previous work because of the parasitic voltage drop inside the electrodes. The miniaturized device was much better in that respect, because the channel and gate resistance remained constant with 
lateral proportional scaling, while the crossover area 
was drastically reduced. This means that the same current density in the dielectric was reached at a much lower total current and hence a much lower voltage drop inside the electrodes. 

\section{Conclusion}
The wet etching processes for SRO and LSMO presented here are optimized to give excellent oxide epitaxy on etched surfaces and are highly selective, leaving photoresist and STO intact. The etch process for SRO also does not attack LSMO, and vice versa. Many oxides that are sufficiently stable in water and diluted acids, including many common substrates, should also not be affected. Using oxidation or reduction at a suitable pH seems to be a good approach to design etchants for other mixed oxides with a transition metal ion in an insoluble oxidation state. 

It is considerably more difficult to create an epitaxial heterostructure with lithographic methods instead of simple shadow masks. More work is required, the layers may degrade and subsequent layers have defects unless the right combination of materials, structuring methods and surface conditioning steps is found. The structures created with lithography can, however, be much smaller, precisely aligned, arbitrary shapes are possible and the layers have a well-defined thickness over the entire device area: see, for example, \cref{fig:STO-AFM} (a). 

We also show in this paper that, as expected, scaling down the devices reduces the problems with parasitic voltage drops and increases the maximum theoretically possible current density. Why the LSMO channel changes its conductance only minimally must be investigated in subsequent experiments. This could be related to the results of Li et al.\cite{Li20071508} who found at intermediate temperature a superior performance of \ce{La2CuO4}-based cathode materials for solid oxide fuel cells compared to LSMO-based cathodes in terms of oxygen exchange and oxygen transport speed.

\section*{Acknowledgements}
We acknowledge Michael Faley and Yuri Divin (PGI-5) for valuable help and discussion, Rolf Speen (PGI-5) for the 3D artwork, and René Borowski and Mirka Grates (PGI-7) for performing ion beam milling and assistance with cleanroom work. We thank our neighboring institute PGI-7 for joint use of instruments and facilities. The software WSXM~\cite{horcas:013705} was used to process the AFM data.

\section*{Conflict of interest statement}
Dieter Weber, Róza Vőfély, Yuehua Chen and Ulrich Poppe have filed a patent application for the etching methods described.

\bibliography{references}

\end{document}